\documentclass[english,aps,manuscript]{revtex4}
\usepackage[T1]{fontenc}
\usepackage[latin1]{inputenc}
\usepackage{babel}
\usepackage{graphics}

\makeatother
\begin{document}

\title{Cooper pairing with finite angular momentum: \\
BCS vs Bose limits}

\author{Jorge Quintanilla}

\email{quintanj@th.ph.bham.ac.uk}

\altaffiliation{Present address: Theoretical Physics Research Group, School of
Physics and Astronomy, University of Birmingham, Edgbaston, Birmingham B15 2TT,
U.K.}

\affiliation{Departamento de F\'{\i}sica e Inform\'atica, Instituto de F\'{\i}sica
de S\~{a}o Carlos, Universidade de S\~{a}o Paulo, Caixa Postal 369,
S\~{a}o Carlos SP 13560-970, Brazil}

\author{Balazs L. Gy\"orffy}

\affiliation{H.H. Wills Physics Laboratory, University of Bristol, Tyndall Av.,
Bristol BS8 1TL, U.K.}

\begin{abstract}
We revisit the old problem of exotic superconductivity as Cooper pairing
with finite angular momentum emerging from a central potential. Based
on some general considerations, we suggest that the phenomenonn is
associated with interactions that keep electrons at some particular,
finite distance \( r_{0} \), and occurs at a range of intermediate
densities \( n\sim 1/r_{0}^{3} \). We discuss the ground state and
the critical temperature in the framework of a standard functional-integral
theory of the BCS to Bose crossover. We find that, due to the lower
energy of two-body bound states with \( l=0 \), the rotational symmetry
of the ground state is always restored on approaching the Bose limit.
Moreover in that limit the critical temperature is always higher for
pairs with \( l=0. \) The breaking of the rotational symmetry of
the continuum by the superfluid state thus seems to be a property
of weakly-attractive, non-monotonic interaction potentials, at intermediate
densities. 
\end{abstract}
\maketitle

\section{Introduction}

Since the original formulation of BCS theory \cite{Bardeen-Cooper-Schrieffer-57,DeGennes-66}
there has been great interest in possible new phenomena arising from
its generalisation. The earliest example concerns the possibility
of Cooper pairs having angular momentum quantum number \( l>0 \),
thus breaking the rotational symmetry of the continuum \cite{Balian-64,Anderson-Brinkman-74}.
Such speculations were based on the assumtion of a different \emph{shape}
of the potential describing the effective attraction between fermions.
Another generalisation concerned \emph{stronger} values of the fermion-fermion
attraction. It was realised that there is a crossover, as this strength
is increased, from a BCS superfluid to a Bose-Einstein condensate
of non-overlapping pairs \cite{Eagles-69}, with a dramatic effect
on the critical temperature \cite{Pincus-et-al-73}. It turned out
that the ground state can be described by a straightforward generalisation
of BCS theory \cite{Leggett-80}, while the superconducting instability
requires taking into account {}``preformed pair'' (PP) fluctuations
in the normal state \cite{Nozieres-SchmittRink-85}. 

Experimentally, Cooper pairing with \( l>0 \) was first observed
in the superfluid state of \( ^{3} \)He. Moreover since the discovery
of superconductivity in cuprate preovskites \cite{Bednorz-Muller-86},
whose pairs have \( d_{x^{2}-y^{2}} \) symmetry \cite{Annett-Goldenfeld-Leggett-96},
several families of {}``anomalous superconductors'' have been found.
Many of these materials present exotic pairing in the form of a superconducting
state that breaks the rotational symmetry of the crystal, and they
often deviate from BCS theory in several other ways \cite{Annett-99}.
This led to a resurgence of interest in the BCS to Bose crossover
\cite{Micnas-Ranninger-Robaskiewicz-90,Gyorffy-Staunton-Stocks-91,Zwerger-92,Alexandrov-Rubin-93,Haussmann-93,Pistolesi-Strinati-94,Haussmann-94,Pistolesi-Strinati-96,SaDeMelo-Randeria-Engelbrecht-93,Engelbrecht-Randeria-SaDeMelo-97,Babaev-Kleinert-98,Babaev-Kleinert-99,Andrenacci-Pieri-Strinati-99,Chen-Kosztin-Levin-00,Ohashi-Griffin-02},
particularly in models with exotic pairing
\cite{Randeria-Duan-Shieh-90,Zwerger-97,Engelbrecht-Nazarenko-Randeria-98,DenHertog-99,Andrenacci-Perali-Pieri-Strinati-99,Jon-James-2000,Soares-Kokubun-RodriguezNunez-Rendon-02,Quintanilla-Gyorffy-Annett-Wallington-02}.
On the other hand little attention has been given to the physics of
the crossover in the context of the original discussions of exotic
pairing \cite{Balian-64,Anderson-Brinkman-74}, namely when a central
attraction, in a continuum, leads to Cooper pairing with \( l>0 \).
In fact this is a specially interesting case since the two-body ground
state is guaranteed, under quite general conditions, \cite{Landau-Lifshitz-QM}
to have \( l=0 \), making exotic pairing necessarily a many-body
effect. Some obvious questions arise: What type of isotropic interactions
lead to exotic pairing, and under what conditions? Is exotic pairing
possible in the Bose-Einstein (BE) limit, when the BCS ground state
is a condensate of non-overlapping pairs \cite{Eagles-69,Pincus-et-al-73,Leggett-80,Nozieres-SchmittRink-85}?
Moreover the recent achievement of Fermi degeneracy in magnetically
trapped, ultra-low temperature gases has stimulated speculations that
superfluidity \cite{Bruun-Castin-Dum-Burnett-99,Combescot-01,Holland-Kokkelmans-Chiofalo-Walser-01}
and indeed the BCS-Bose crossover \cite{Ohashi-Griffin-02} may be
observable in these systems. Understanding the above questions may
guide us as to whether exotic pairing is also a possibility.

Recently, we and our co-authors have studied the above questions in
a simple model that features exotic pairing via a central attraction
\cite{Quintanilla-Gyorffy-Annett-Wallington-02}. In this contribution
we revisit them with a more general point of view, using some of the
results obtained for the {}``delta shell model'' (DSM) of
Ref.~\cite{Quintanilla-Gyorffy-Annett-Wallington-02}
as an illustration. We begin section \ref{SEC-central potentials leading to exotic pairing}
by reminding the reader how, as a consequence of the weak-coupling
theory of superconductivity, a central potential can, \emph{in principle},
lead to exotic pairing \cite{Balian-64,Anderson-Brinkman-74}. Then
we take this old argument one little step further by asking: what
is the essential feature that makes a particular interaction potential,
\emph{in practice}, lead to this effect? Having established the existence
of such potentials, and knowing what they look like, we move on to
outline, in section \ref{SEC-basic theory}, the main features of
a simple, but fairly general theory of the BCS to Bose crossover.
Our formalism follows a recipe that has, by now, become standard \cite{Randeria-95}
(though not the only one \cite{Gyorffy-Staunton-Stocks-91,Chen-Kosztin-Levin-00}):
to introduce bosonic pairing fields via a Hubbard-Stratonovich transformation
(HST) \cite{Bender-Orszag-78,Nagaosa-99} and then expand the action
to quadratic (Gaussian) order in those fields. Although limited \cite{Haussmann-93,Haussmann-94},
such scheme suffices for the discussion of the BCS and BE limits.
The novelty of our formulation is that pairs may be created and annihilated
with different values of their angular momentum quantum numbers. In
contrast, many previous applications of the Gaussian theory assumed
these internal degrees of freedom of the Cooper pairs to be {}``frozen''
to the desired \( s \) \cite{Zwerger-92,SaDeMelo-Randeria-Engelbrecht-93}
or \( d_{x^{2}-y^{2}} \) \cite{Zwerger-97} state. As we shall see
these internal degrees of freedom turn out to be important, determining
some of the key features of the problem. In sections \ref{SEC-gs}
and \ref{SEC-Tc} we apply the formalism to discuss the ground state
and the superconducting instability, respectively, in the BE limit,
where the original argument for exotic pairing cannot be applied.
Finally, in section \ref{SEC-conclusion}, we offer our conclusions.

\section{Central potentials leading to exotic pairing\label{SEC-central potentials leading to exotic pairing}}

Let us begin by recalling some central ideas of the original weak-coupling
 theory  of exotic pairing \cite{Balian-64,Anderson-Brinkman-74}.
Consider a system of electrons, in a three-dimensional continuum,
interacting via some local, non-retarded, central potential \( V\left( \mathbf{r}-\mathbf{r}'\right) =V\left( \left| \mathbf{r}-\mathbf{r}'\right| \right)  \).
For simplicity we will assume the interaction to take place between
electrons with opposite spins and limit our discussion to the case
of singlet pairing, with angular momentum quantum number \( l=0,2,4,\ldots  \)
As is well-known the Fourier transform \begin{equation}
\label{FT of V inverted opposite spins}
V\left( \mathbf{k}-\mathbf{k}'\right) =\int d^{3}\mathbf{r}e^{i\left( \mathbf{k}-\mathbf{k}'\right) .\mathbf{r}}V\left( \mathbf{r}\right) 
\end{equation}
admits the partial-wave decomposition

\begin{equation}
\label{angle decomp of V}
V\left( \mathbf{k}-\mathbf{k}'\right) =\sum _{l=0}^{\infty }K_{l}\left( \left| \mathbf{k}\right| ,\left| \mathbf{k}'\right| \right) \left( 2l+1\right) P_{l}\left( \hat{\mathbf{k}}.\hat{\mathbf{k}}'\right) 
\end{equation}
in terms of the Legendre polynomials \( P_{l}\left( \hat{\mathbf{k}}.\hat{\mathbf{k}}'\right) =\left( 2l+1\right) ^{-1}4\pi \sum _{m}Y_{l,m}\left( \hat{\mathbf{k}}\right) Y_{l,m}^{*}\left( \hat{\mathbf{k}}'\right)  \).
It was soon pointed out \cite{Balian-64} that in the weak-coupling
limit one can use the approximation\begin{equation}
\label{angle decomp of V with wc approx}
V\left( \mathbf{k}-\mathbf{k}'\right) \approx K_{l_{max}}\left( 2l_{max}+1\right) P_{l_{max}}\left( \hat{\mathbf{k}}.\hat{\mathbf{k}}'\right) ,
\end{equation}
where \( l_{max} \) is the value of \( l \) for which the coupling
constant on the Fermi surface, \begin{equation}
\label{BCS limit coupling const}
K_{l}\equiv K_{l}\left( k_{F},k_{F}\right) ,
\end{equation}
 is largest (\( k_{F} \) is the Fermi vector). The approximate form
(\ref{angle decomp of V with wc approx}) of the potential \( V\left( \mathbf{k}-\mathbf{k}'\right)  \)
is, for \( l_{max}>0 \), anisotropic, and it leads to pairing with
finite angular momentum quantum number \( l_{max} \) \cite{Balian-64}. 

\begin{figure}
{\centering \resizebox*{0.5\textwidth}{!}{\includegraphics{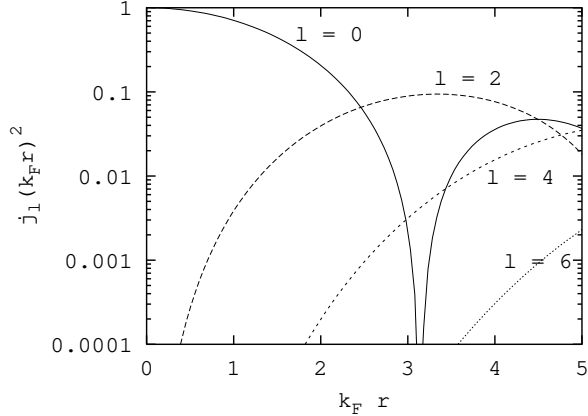}} \par}

\caption{\label{FIG-weighting factor}The weighting factors in the integral
on the right-hand side of Eq. (\ref{Kl}) for pairing with the first
four even values of the angular momentum quantum number, \protect\( l=0,2,4,6\protect \).
Note the logarithmic scale on the vertical axis.}
\end{figure}

\begin{figure}
{\centering \resizebox*{0.7\textwidth}{!}{\includegraphics{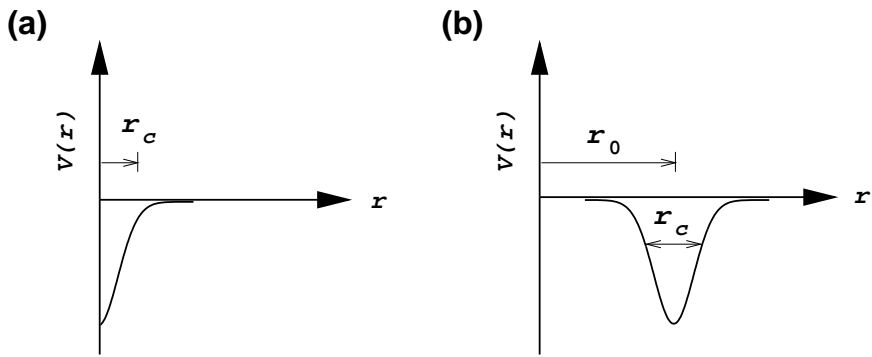}} \par}

\caption{\label{FIG-two_plots}Interaction potentials representing on-site
attraction (a) and attraction at a finite distance (b). }
\end{figure}

Let us now try to understand how the preference for a particular value
of \( l \) comes about. To do this one has to examine the relationship
between the coupling constant \( K_{l} \) and the interaction potential
\( V\left( r\right)  \). It can be found by expanding in spherical
waves the two exponentials \( e^{i\mathbf{k}.\mathbf{r}} \) and \( e^{-i\mathbf{k}'.\mathbf{r}} \)
in Eq.~(\ref{FT of V inverted opposite spins}); comparison to (\ref{angle decomp of V})
yields \begin{equation}
\label{kernel in general}
K_{l}\left( \left| \mathbf{k}\right| ,\left| \mathbf{k}'\right| \right) =\left( -1\right) ^{l}\int _{0}^{\infty }dr\, 4\pi r^{2}\, j_{l}\left( \left| \mathbf{k}\right| r\right) \, V\left( r\right) \, j_{l}\left( \left| \mathbf{k}'\right| r\right) ,
\end{equation}
where \( j_{l}\left( x\right)  \) is a spherical Bessel function.
Substituing (\ref{kernel in general}) in (\ref{BCS limit coupling const})
we obtain\begin{equation}
\label{Kl}
K_{l}=\left( -1\right) ^{l}\int _{0}^{\infty }dr\, 4\pi r^{2}\, j_{l}\left( k_{F}r\right) ^{2}\, V\left( r\right) .
\end{equation}
Thus \( K_{l} \) is a weighted integral of the interaction potential
\( V\left( r\right)  \). The weighting factors \( j_{l}\left( k_{F}r\right) ^{2} \)
are shown in Fig.~\ref{FIG-weighting factor} for \( l=0,2,4,\ldots  \)
Evidently for a purely attractive, monotonic potential, such as the
one in Fig.~\ref{FIG-two_plots}~(a), \( l_{max}=0 \). To have
\( l_{max}>0 \), \( V\left( r\right)  \) has to be most attractive
at distances at which the weighting factors for a finite value of
the angular momentum quantum number are largest. That is achieved
by non-monotonic potentials that lead to maximum attraction at some
finite distance \( r_{0} \), such as the one in Fig.~\ref{FIG-two_plots}~(b).
For example, provided that the width of the potential well centred
on \( r_{0} \), namely \( r_{c} \) (see figure), is sufficiently
small, choosing \( r_{0} \) so that \( k_{F}r_{0}\sim 3 \) yields
\( l_{max}=2 \). 

\section{Basic theory of the BCS to Bose crossover\label{SEC-basic theory}}

The above discussion implies that, in the BCS limit, exotic pairing
can be described by an effectively anisotropic interaction, Eq.~(\ref{angle decomp of V with wc approx}).
On the other hand we would like to describe the ground state and the
superconducting instability also in the BE limit, where such approximation
may not (and, as we shall see, does not) apply. For this we need to
develop our theory in a more general framework, and in particular
it is important to work with a complete description of the interaction
potential. For concreteness let us write down explicitely the grand-canonical
Hamiltonian for the situation at hand. At chemical potential \( \mu  \),
it is \cite[Eq. (1) (for example)]{Nozieres-SchmittRink-85}\begin{eqnarray}
\hat{H}-\mu \hat{N} & = & \sum _{\mathbf{k},\sigma }\varepsilon _{\mathbf{k}}\hat{c}_{\mathbf{k},\sigma }^{+}\hat{c}_{\mathbf{k},\sigma }+\frac{1}{L^{3}}\sum _{\mathbf{q},\mathbf{k},\mathbf{k}'}V\left( \mathbf{k}-\mathbf{k}'\right) \times \nonumber \\
 &  & \hspace {2cm}\times \hat{c}^{+}_{\mathbf{q}/2+\mathbf{k},\uparrow }\hat{c}^{+}_{\mathbf{q}/2-\mathbf{k},\downarrow }\hat{c}_{\mathbf{q}/2-\mathbf{k}',\downarrow }\hat{c}_{\mathbf{q}/2+\mathbf{k}',\uparrow },\label{our H in k-space} 
\end{eqnarray}
 It is particularly illustrative to re-write it in the following way:\begin{eqnarray}
\hat{H}-\mu \hat{N} & = & \sum _{\mathbf{k},\sigma }\varepsilon _{\mathbf{k}}\hat{c}_{\mathbf{k},\sigma }^{+}\hat{c}_{\mathbf{k},\sigma }+\frac{1}{L^{3}}\sum _{\kappa ,l,m}\sum _{\mathbf{q}}V_{\kappa ,l}\hat{b}^{+}_{\kappa ,l,m,\mathbf{q}}\hat{b}_{\kappa ,l,m,\mathbf{q}}.\label{our H in k-space separated} 
\end{eqnarray}
Here, \( \hat{c}^{+}_{\mathbf{k},\sigma },\hat{c}_{\mathbf{k},\sigma } \)
create and annihilate, respectively, and electron with momentum \( \hbar \mathbf{k} \)
and spin \( \sigma =\uparrow ,\downarrow  \). \( \varepsilon _{\mathbf{k}}\equiv \hbar ^{2}\left| \mathbf{k}\right| ^{2}/2m^{*}-\mu  \)
is the single-particle dispersion relation (\( m^{*} \) is the effective
mass of an electron) and \( L^{3} \) is the (very large) sample volume.
The additional operators\begin{eqnarray}
\hat{b}^{+}_{\kappa ,l,m,\mathbf{q}} & \equiv  & \sum _{\mathbf{k}}\phi _{\kappa ,l,m,\mathbf{k}}\hat{c}^{+}_{\mathbf{q}/2+\mathbf{k}\uparrow }\hat{c}^{+}_{\mathbf{q}/2-\mathbf{k}\downarrow },\label{b+ operator} \\
\hat{b}_{\kappa ,l,m,\mathbf{q}} & \equiv  & \sum _{\mathbf{k}}\phi _{\kappa ,l,m,\mathbf{k}}^{*}\hat{c}_{\mathbf{q}/2-\mathbf{k}\downarrow }\hat{c}_{\mathbf{q}/2+\mathbf{k}\uparrow }.\label{b operator} 
\end{eqnarray}
create and annihilate, respectively, a pair with opposite spins, total
momentum \textbf{\( \hbar \mathbf{q} \)} and internal wave function
\( \phi _{\kappa ,l,m,\mathbf{k}}\equiv R_{\kappa ,l}\left( \left| \mathbf{k}\right| \right) Y_{l,m}\left( \hat{\mathbf{k}}\right)  \),
where \( \hbar \mathbf{k} \) is the momentum of one of the two components
of the pair with respect to its centre of mass. To put the generic
Hamiltonian (\ref{our H in k-space}) in the form (\ref{our H in k-space separated})
it suffices to define the {}``kernel factors'' \( R_{\kappa ,l}\left( \left| \mathbf{k}\right| \right)  \)
and {}``coupling constants'' \( V_{\kappa ,l} \) so that they yield
the following parametrization:\begin{equation}
\label{central V amplitude in terms of its eigenfunctions}
K_{l}\left( \left| \mathbf{k}\right| ,\left| \mathbf{k}'\right| \right) =\frac{1}{4\pi }\sum _{\kappa }V_{\kappa ,l}R_{\kappa ,l}\left( \left| \mathbf{k}\right| \right) R^{*}_{\kappa ,l}\left( \left| \mathbf{k}'\right| \right) .
\end{equation}
We may construct the \( \phi _{\kappa ,l,m,\mathbf{k}} \) and \( V_{\kappa ,l} \)
as a complete set of solutions to the eigenvalue problem\begin{equation}
\label{V eigenvalue eq}
\sum _{\mathbf{k}'}V\left( \mathbf{k}-\mathbf{k}'\right) \phi _{\kappa ,l,m,\mathbf{k}'}=V_{\kappa ,l}\phi _{\kappa ,l,m,\mathbf{k}},
\end{equation}
which can be regarded as the result of neglecting, for strong attraction,
the kinetic energy contribution to the Schr\"odinger equation for
a two-body bound state. 

Eq.~(\ref{our H in k-space separated}) generalises the usual BCS
Hamiltonian by including interaction terms corresponding to pairs
with \( \hbar \mathbf{q}\neq 0 \) and different values of the the
internal angular momentum, given by \( l,m \). This is required for
the correct description of the normal state above \( T_{c} \) in
the strong-coupling limit \cite{Nozieres-SchmittRink-85} and, as
we shall see, to capture the essential physics of exotic pairing via
a central potential, respectively. The later can be understood, in
essence, by noting that the angular momentum quantum numbers \( l,m \)
describe the internal rotational degrees of freedom of the Cooper
pairs, and so they are a key ingredient to describe their dynamics
in the case of a central attraction. The additional index \( \kappa  \)
has been introduced to ensure the generality of (\ref{central V amplitude in terms of its eigenfunctions}),
and it refers to the relative motion of the electrons in a pair in
the radial direction. Evidently for interaction potentials of the
type we are interested in such radial motion is {}``locked'' so
that the distance between the two electrons remains constant, and
equal to \( r_{0} \). We will therefore disregard these vibrational
modes, assuming that there is a single kernel factor \( R_{l}\left( \left| \mathbf{k}\right| \right)  \)
for each value of \( l \). Obviously in that case there is a single
coupling constant \( V_{l} \) for each value of the angular momentum
quantum number. This approximation becomes exact in the limit when
\( r_{c}\rightarrow 0 \) and \( V\left( r_{0}\right) \rightarrow -\infty  \),
keeping \( V\left( r_{0}\right) r_{c}\equiv -g \) constant. Then
we obtain the DSM \cite{Quintanilla-Gyorffy-00}, featuring the central
{}``delta shell'' potential \cite{Gottfried-66}\begin{equation}
\label{delta-shell potential}
V\left( \mathbf{r}-\mathbf{r}'\right) =-g\delta \left( \left| \mathbf{r}-\mathbf{r}'\right| -r_{0}\right) ,
\end{equation}
for which \cite{Quintanilla-Gyorffy-Annett-Wallington-02} \begin{equation}
\label{V separable DSM}
K_{l}\left( \left| \mathbf{k}\right| ,\left| \mathbf{k}'\right| \right) =-g4\pi r_{0}^{2}\left( -1\right) ^{l}j_{l}\left( \left| \mathbf{k}\right| r_{0}\right) j_{l}\left( \left| \mathbf{k}'\right| r_{0}\right) .
\end{equation}
More generally, it must be regarded as a convenient approximation,
valid when the potential well in Fig.~\ref{FIG-two_plots}~(b) is
sufficiently deep and narrow.

Having set up our Hamiltonian, we can now use the standard method
reviewed by Randeria \cite{Randeria-95}, which has been applied to
specific models by several authors \cite{Zwerger-92,SaDeMelo-Randeria-Engelbrecht-93,Zwerger-97,Engelbrecht-Randeria-SaDeMelo-97},
to discuss the ground state at \( T=0 \) and the superconducting
instability at \( T_{c} \). In short we introduce bosonic fields
\( \Delta _{l,m,\mathbf{q}}\left( \tau \right) ,\Delta ^{*}_{l,m,\mathbf{q}}\left( \tau \right)  \)
coupling to the pair creation and annihilation operators \( \hat{b}^{+}_{l,m,\mathbf{q}},\hat{b}_{l,m,\mathbf{q}} \),
respectively. In terms of these fields one can define an effective
action \( S_{eff}\left[ \Delta ^{*},\Delta \right]  \) that determines
the partition function of the system: \begin{equation}
\label{Z with Seff}
Z=\int D\left[ \Delta ^{*},\Delta \right] \, e^{-S_{eff}\left[ \Delta ^{*},\Delta \right] }.
\end{equation}
 We then expand it to the lowest non-trivial (Gaussian) order: \begin{equation}
\label{Seff expansion}
S_{eff}\left[ \Delta ^{*},\Delta \right] \approx S_{eff}\left[ \Delta ^{*(0)},\Delta ^{(0)}\right] +S^{(2)}_{eff}\left[ \Delta ^{*},\Delta \right] .
\end{equation}
Here \( \Delta _{l,m,\mathbf{q}}^{*(0)}\left( \tau \right) ,\Delta _{l,m,\mathbf{q}}^{(0)}\left( \tau \right)  \)
is the configuration of the pairing fields at the saddle point and
the Gaussian contribution, \( S_{eff}^{(2)}\left[ \Delta ^{*},\Delta \right]  \),
takes into account fluctuations of the fields about that saddle point.

\section{Exotic pairs in the ground state \label{SEC-gs}}

In the ground state, the pairing fields become {}``frozen'' in their
saddle-point configurations \cite{Randeria-95}. \( S_{eff}^{(2)}\left[ \Delta ^{*},\Delta \right]  \)
acquires a trivial form such that \( \int D\left[ \Delta ^{*},\Delta \right] \exp \left\{ -S_{eff}^{(2)}\left[ \Delta ^{*},\Delta \right] \right\} =1 \),
Eq.~(\ref{Z with Seff}) thus becoming \( Z=\exp \left\{ -S_{eff}\left[ \Delta ^{*(0)},\Delta ^{(0)}\right] \right\}  \).
As usual we look for a stationary and homogeneous saddle point:\begin{eqnarray*}
\Delta _{l,m,\mathbf{q}}^{(0)}\left( \tau \right)  & = & \Delta _{l,m}^{(0)}\delta _{\mathbf{q},0},\\
\Delta _{l,m,\mathbf{q}}^{*(0)}\left( \tau \right)  & = & \Delta _{l,m}^{(0)*}\delta _{\mathbf{q},0},
\end{eqnarray*}
which yields the BCS ground state \cite{DeGennes-66}. As is well-known
this state can describe, at least variationally, the BCS to Bose crossover
at \( T=0 \) \cite{Leggett-80,Nozieres-SchmittRink-85,Andrenacci-Perali-Pieri-Strinati-99}.
The amplitudes \( \Delta _{l,m}^{(0)} \) are related to the expectation
values of the pair annihilation operators by \( \Delta _{l,m}^{(0)}=V_{l}\left\langle \hat{b}_{l,m,0}\right\rangle  \).
The equations determining the saddle point take the following form:\begin{eqnarray}
\Delta _{l,m}^{(0)} & = & -V_{l}\sum _{l',m'}\left\{ \int \frac{d^{3}\mathbf{k}}{\left( 2\pi \right) ^{3}}\frac{\phi _{l,m,\mathbf{k}}^{*}\phi _{l',m',\mathbf{k}}}{2E_{\mathbf{k}}}\right\} \Delta _{l',m'}^{(0)}\label{gap eq DL} 
\end{eqnarray}
where \( E_{\mathbf{k}}\equiv \sqrt{\varepsilon _{\mathbf{k}}^{2}+\left| \Delta _{\mathbf{k}}\right| ^{2}} \)
with \( \Delta _{\mathbf{k}}\equiv \sum _{l,m}\Delta ^{(0)}_{l,m}R_{l}\left( \left| \mathbf{k}\right| \right) Y_{l,m}\left( \hat{\mathbf{k}}\right)  \)
the usual BCS {}``gap function''. Evidently this non-linear system
of equations may have many different solutions, each corresponding
to a different superposition of angular momentum states. To determine
their relative stability one has to evaluate the appropriate potential.
If, as usual, we fix the density \( n \), treating the chemical potential
\( \mu  \) as a parameter to be determined self-consistenly \cite{Nozieres-SchmittRink-85},
the energy \( \mathcal{U}_{0}=\left\langle \hat{H}\right\rangle  \)
has to be calculated. Quite generally, it is given by\begin{equation}
\label{gs energy eigen V}
\frac{1}{L^{3}}\mathcal{U}_{0}=\int \frac{d^{3}\mathbf{k}}{(2\pi )^{3}}\frac{\hbar ^{2}\left| \mathbf{k}\right| ^{2}}{2m^{*}}\left( 1-\frac{\varepsilon _{\mathbf{k}}}{E_{\mathbf{k}}}\right) +\sum _{l,m}\frac{\left| \Delta ^{(0)}_{l,m}\right| ^{2}}{V_{l}}
\end{equation}
which results from taking the \( T\rightarrow 0 \) limit of \( L^{-3}\left\langle \hat{H}\right\rangle =-L^{-3}\beta ^{-1}\ln Z+\mu n \). 

The above equations are entirely equivalent to the usual mean-field
theory \cite{Ketterson-Song-99} for a sufficiently general interaction
potential (at least, when only pairing correlations are taken into
account). In particular, in the limit of very weak attraction (BCS
limit), the argument of section \ref{SEC-central potentials leading to exotic pairing}
applies. Thus we expect that, for the type of interaction that we
are interested in, represented in Fig.~\ref{FIG-two_plots}~(b),
pairing with \( l=2 \) will be preferred to \( l=0 \) for a range
of values of the density, \( n=k_{F}^{3}/3\pi ^{2} \), such that
\( k_{F}r_{0}\sim 3 \). In the extreme case in which the attraction
takes place only exactly at \( r=r_{0} \) one expects the densities
at which the preferred value \( l_{max} \) of the angular momentum
quantum number changes to be given by \begin{equation}
\label{phase transition wc}
j_{0}\left( k_{F}r_{0}\right) ^{2}=j_{2}\left( k_{F}r_{0}\right) ^{2}
\end{equation}
This result, which is degenerate in the quantum number \( m \), is
exact for the DSM (\ref{delta-shell potential}), and it gives the
boundaries of a \emph{quantum phase transition} in which the ground
state changes the rotational symmetry. 

On the other hand in the dilute, strong coupling limit (BE limit),
there is no longer a well-defined Fermi surface and the approximation
(\ref{angle decomp of V with wc approx}) ceases to be useful. The
gap equation (\ref{gap eq DL}) describes a two-body bound state with
energy \( \varepsilon _{b}^{l}=2\mu  \) and wave function \( \psi _{\mathbf{k}}=\Delta _{\mathbf{k}}/2E_{\mathbf{k}} \)
\cite{Leggett-80,Nozieres-SchmittRink-85,Randeria-Duan-Shieh-90}
and so, evidently, the full functional form of the interaction potential
\( V\left( \mathbf{r}-\mathbf{r}'\right)  \) has to be taken into
account. Since \( l \) is a good quantum number for the Schr\"odinger equation,
a well-defined angular momentum is a shared characteristic of the
BCS and BE limits. On the other hand it is easy to show, using \( \mu \ll -\left| \Delta _{l,m}^{(0)}\right|  \),
that in the BE limit Eq.~(\ref{gs energy eigen V}) becomes \begin{equation}
\label{sc-energy}
\frac{1}{L^{3}}\mathcal{U}_{0}=\frac{1}{2}n\varepsilon _{b}^{l}.
\end{equation}
This equation has a simple interpretation: in the dilute, strong-coupling
limit, the system is a BE condensate of non-overlapping pairs. Each
pair has \( \hbar \mathbf{q}=0 \) and they are all in the same internal
state with energy \( \varepsilon _{b}^{l} \). It follows that the
energy of the system is given simply by the formation energies of
the individual pairs. Since, for a central potential, the bound state
with lowest energy always has \( l=0 \), one expects that \emph{rotational
symmetry is never broken in the BE limit}.

\begin{figure}
{\centering \resizebox*{0.5\textwidth}{!}{\includegraphics{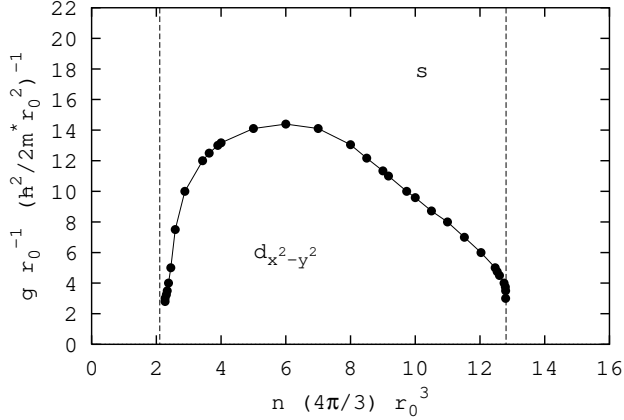}} \par}

\caption{\label{FIG-relative-stability}Phase diagram of the relative stability
of the \protect\( s\protect \) and \protect\( d_{x^{2-y^{2}}}\protect \)
trial ground states of the delta-shell model, taken from Ref.~\cite{Quintanilla-Gyorffy-Annett-Wallington-02}.
In the limit of small coupling constant \protect\( g\protect \),
the phase transition takes place precisely at the values of the density
\protect\( n=k_{F}^{3}/3\pi ^{2}\protect \) for which the condition
(\ref{phase transition wc}) holds. These are marked by the dashed
vertical lines.}
\end{figure}

Fig.~\ref{FIG-relative-stability} shows a specific result for the
DSM \cite{Quintanilla-Gyorffy-Annett-Wallington-02} that illustrates
the above points. It is a phase diagram for the relative stability
of ground states in which Cooper pairs have \( s \) and \( d_{x^{2}-y^{2}} \)
symmetries. At weak coupling, pairing with \( l=0 \) is preferred
at low and high densities, with a quantum phase transition leading
to the onset of \( d_{x^{2}-y^{2}} \) pairing in the intermediate-density
regime. The location of the phase boundary is accurately predicted
by Eq.~(\ref{phase transition wc}). In contrast, as the attraction
is made stronger the range of densities over which the state with
\( l=2 \) is preferred becomes narrower until, above some critical
value of the coupling constant, rotational symmetry is restored for
all values of the density. The \( d_{x^{2}-y^{2}} \) state is thus
confined to a relatively small {}``island'' in parameter space.
This result not only confirms our expectation that pairing should
take place in the \( s \) state in the BE limit, but in fact suggests
that \( l=0 \) is preferred at all densities, provided that the attraction
is sufficiently strong. However note that the phase boundary, at finite
values of the coupling constant \( g \), is not necessarily degenerate
in \( m \) (unlike at weak coupling). Thus investigations carried
out with more general trial ground states (allowing, for example,
for mixing of \( s \) and \( d \) symmetries) have to be carried
out.

\section{Superconducting critical temperature for exotic pairs\label{SEC-Tc}}

When the attraction is strong, fluctuations around the saddle-point
configurations of the fields are important to determine the low-lying
excitations of the BCS ground state \cite{Alexandrov-Rubin-93,Engelbrecht-Randeria-SaDeMelo-97}
as well as to describe the superconducting instability at \( T_{c} \)
\cite{Zwerger-92,SaDeMelo-Randeria-Engelbrecht-93,Zwerger-97}. Our
interest here is in the later, as we wish to see whether such instability
can correspond, in the strong-coupling, low-density limit, to the
formation of a BE condensate of \emph{exotic} pairs. 

Just above \( T_{c} \), the partition function (\ref{Z with Seff})
is given by \begin{equation}
\label{Z equals Z0 times dZ}
Z=Z_{0}\times \delta Z
\end{equation}
 where \( Z_{0}=\exp \left\{ -S\left[ 0,0\right] \right\}  \) corresponds
to a free electron gas while \( \delta Z=\int D\left[ \Delta ^{*},\Delta \right] \exp \left\{ S^{(2)}_{eff}\left[ \Delta ^{*},\Delta \right] \right\}  \)
is the contribution from PP just above \( T_{c} \). The normal state
is composed of a mixture of two gases, one made out of free electrons
and the other one consisting of PP. In effect, on account of (\ref{Z equals Z0 times dZ})
the total electron density \( n\left( \beta ,\mu \right) =L^{-3}k_{B}T\partial \ln Z/\partial \mu  \)
can be written as\begin{equation}
\label{gaussian-density-eq}
n\left( \beta ,\mu \right) =n_{0}\left( \beta ,\mu \right) +\delta n\left( \beta ,\mu \right) 
\end{equation}
where the first and second terms on the right hand side come from
\( Z_{0} \) and \( \delta Z \), respectively. In particular the
contribution from free electrons has the familiar form \( n_{0}\left( \beta ,\mu \right) =2L^{-3}\sum _{\mathbf{k}}f\left( \beta \varepsilon _{\mathbf{k}}\right)  \)
where \( f\left( x\right) \equiv \left( e^{x}+1\right) ^{-1} \) is
the Fermi distribution function. As is well known in the BCS limit
\( \mu \beta \rightarrow \infty  \) all electrons are unpaired just
above \( T_{c} \) so we have \( n\approx n_{0} \). Conversely, in
a dilute system with strong attraction (BE limit) we have \( n\approx \delta n \):
the normal state just above \( T_{c} \) is composed exclusively of
preformed pairs, whose BE condensation leads to superconductivity. 

\begin{figure}
{\centering \resizebox*{0.45\textwidth}{!}{\includegraphics{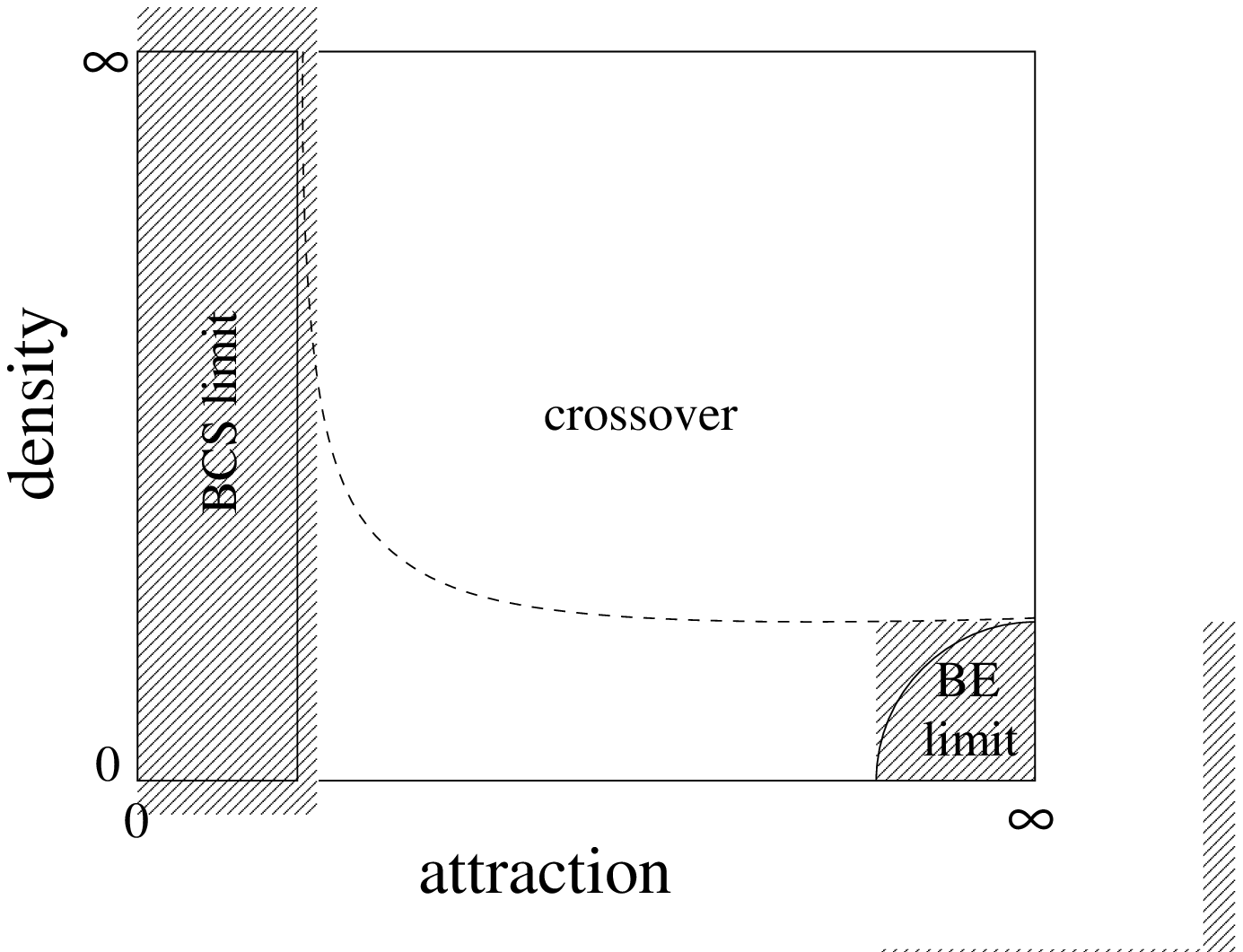}} \par}

\caption{\label{FIG-validity}Validity of the Gaussian description of the
superconducting instability \cite{Randeria-95}. The theory is deemed
correct in the shaded areas: BCS limit (weak-coupling) and BE limit
(strong-coupling \emph{and} low densities). For the crossover region
it can be regarded as a convenient interpolation scheme at low densities
(below the dashed line), while at larger densities it fails (see text).}
\end{figure}

It is important to note that the above functional-integral theory
is only valid either at weak coupling or for sufficiently dilute systems
with strong attraction. The problem is (leaving aside the fact that
the it relies on an expansion in powers of the \emph{amplitude} of
the pairing fields, while neglecting other fluctuations due exclusively
to their \emph{phase}, which are essential in two dimensions \cite{Loktev-Quick-Sharapov-01})
that as the density is increased interactions between the PP, which
the Gaussian theory neglects, become important. Similarly, when the
interaction becomes weaker the PP increase their radius and begin
to overlap. Quite generally, at low densities and intermediate coupling
the Gaussian theory of \( T_{c} \) must be regarded only as a convenient
interpolation scheme \cite{Nozieres-SchmittRink-85,Randeria-95},
while it fails completely at higher densities (as is evidenced for
example in the negative bosonic mass obtained for the DSM at sufficiently
large values of the density and the coupling constant \cite{Quintanilla-Gyorffy-Annett-Wallington-02}).
Such limitations (illustrated in Fig.~\ref{FIG-validity}) have been,
at least partially, addressed \cite{Haussmann-93,Haussmann-94,Pieri-Strinati-00},
however for simplicity our discussion of \( T_{c} \) below refers
only to the regions of parameter space in which the standard theory
\cite{Randeria-95} can be safely applied at that temperature. These
are illustrated in Fig.~\ref{FIG-validity}.

Let us now see what form the above standard equations \cite{Randeria-95}
take in the case of our fairly general model. In terms of the bosonic
Matsubara frequencies \( \omega _{\nu }=2\nu \pi /\beta  \) the quadratic
contribution to the effective action has the form \begin{equation}
\label{Sb}
S_{eff}^{(2)}\left[ \Delta ^{*},\Delta \right] =\beta \sum _{\mathbf{q}\nu }\sum _{l,m,l',m'}\Delta ^{*}_{l,m,\mathbf{q}}\left( \omega _{\nu }\right) \Gamma ^{-1}_{l,m,l',m'}\left( \mathbf{q},i\omega _{\nu }\right) \Delta _{l',m',\mathbf{q}}\left( \omega _{\nu }\right) 
\end{equation}
 where the sums on \( l \) and \( l' \) extend only over values
of the angular momentum quantum number having the same parity (both
even or both odd). As usual, we start by writing out the inverse bosonic
propagator explicitly. It is\begin{eqnarray}
\Gamma ^{-1}_{l,m,l',m'}\left( \mathbf{q},i\omega _{\nu }\right)  & = & -\frac{L^{3}}{V_{l}}\delta _{l,l'}\delta _{m,m'}-\sum _{\mathbf{k}}i^{l'-l}\phi _{l,m,\mathbf{k}}^{*}\phi _{l',m',\mathbf{k}}\times \nonumber \\
 &  & \hspace {-2cm}\times \frac{1}{\beta }\sum _{n}G_{0}\left( \frac{\mathbf{q}}{2}+\mathbf{k},i\omega _{n}\right) G_{0}\left( \frac{\mathbf{q}}{2}-\mathbf{k},i\omega _{\nu }-i\omega _{n}\right) \label{inv Gamma} 
\end{eqnarray}
where \( G_{0}\left( \mathbf{k},i\omega _{n}\right) \equiv \left( \varepsilon _{\mathbf{k}}-i\omega _{n}\right) ^{-1} \)
is a free-electron Green's function and \( \omega _{n}=(2n+1)\pi /\beta  \)
a fermion Matsubara frequency. This is a slightly more general form
of the similar expression found in the literature
\cite{Nozieres-SchmittRink-85,Zwerger-92,SaDeMelo-Randeria-Engelbrecht-93,Zwerger-97,Quintanilla-Gyorffy-Annett-Wallington-02}.
Such fairly general bosonic propagator, like the one obtained in Ref.~\cite{Quintanilla-Gyorffy-Annett-Wallington-02}
for the DSM, describes not only the centre-of-mass motion of the PP
existing above \( T_{c} \) (represented by their total momentum \( \hbar \mathbf{q} \))
but also the freedom that they have to change their angular momentum
(represented by the labels \( l,m \)). On the other hand in Refs.~\cite{Nozieres-SchmittRink-85,Zwerger-92,SaDeMelo-Randeria-Engelbrecht-93,Zwerger-97}
these internal degrees of freedom were not taken into account. In
effect, the interaction potentials employed in Ref.~\cite{Nozieres-SchmittRink-85,Zwerger-92,SaDeMelo-Randeria-Engelbrecht-93}
had the form (\ref{angle decomp of V with wc approx}) with \( l_{max}=0 \)
and, therefore, could only lead to pairing in the \( s \) state;
similarly, the potential in Ref.~\cite{Zwerger-97} was chosen so
that it could only lead to pairing in a particular \( d \)-wave state,
with \( d_{x^{2}-y^{2}} \) symmetry. But for strong \emph{central}
attraction one expects that, just above \( T_{c} \), preformed pairs
exist with all values of the angular momentum quantum number, as reflected
by Eq.~(\ref{inv Gamma}). The question that we are trying to answer
here is what pairs will form a BE condensate at \( T_{c} \); in particular,
whether they can have \( l>0 \).

As usual the {}``\( T_{c} \) equation'' is found as the temperature
at which the system becomes unstable with respect to pairing fluctuations
describing a homogeneous, static field:\begin{equation}
\label{instability condition}
\beta \sum _{l,m,l',m'}\Delta ^{*}_{l,m,0}\left( 0\right) \Gamma^{-1}_{l,m,l',m'}\left( 0,0\right) \Delta _{l',m',0}\left( 0\right) =0.
\end{equation}
We obtain \begin{equation}
\label{Tc eq}
1=-\frac{V_{l}}{\left( 2\pi \right) ^{3}}\int _{0}^{\infty }d\left| \mathbf{k}\right| \left| \mathbf{k}\right| ^{2}\left| R_{l}\left( \left| \mathbf{k}\right| \right) \right| ^{2}\frac{1-2f\left( \beta _{c}\varepsilon _{\mathbf{k}}\right) }{2\varepsilon _{\mathbf{k}}}
\end{equation}
which is diagonal in \( l \) and degenerate in \( m \). In general
(\ref{Tc eq}) has several solutions \( \beta _{c}\equiv 1/k_{B}T_{c} \),
corresponding to the formation of a superconducting state with different
values of the angular momentum quantum number \( l=0,2,4,\ldots  \)
Evidently the highest \( T_{c} \) corresponds to the true superconducting
instability, and it gives the angular momentum quantum number of the
Cooper pairs in the superconducting state, just below \( T_{c} \). 

As expected \cite{Randeria-95} Eq.~(\ref{Tc eq}) has the same form
as in the mean-field theory and so the weak-coupling limit the Gaussian
theory reduces to it. In particular the argument for exotic pairing
in the BCS limit that we recalled above applies also to the critical
temperature: thus at intermediate densities we expect pairs to form
and condense simultaneously, at \( T_{c} \), with angular momentum
quantum number \( l_{max}=2 \).

Let us now focus on the BE limit. For sufficiently strong coupling
and low densities the inverse propagator \( \Gamma ^{-1}_{l,m,l',m'}\left( \mathbf{q},i\omega _{\nu }\right)  \)
can be expanded to the lowest non-trivial order in the pair's total
momentum \( \hbar \mathbf{q} \) and frequency \( \omega _{\nu } \)
\cite{Nozieres-SchmittRink-85}. Such expansion was carried out in
Ref.~\cite{Quintanilla-Gyorffy-Annett-Wallington-02} for the DSM,
following closely the procedure of \cite{Zwerger-92,Zwerger-97}.
The derivation is entirely analogous in the present slightly more
general case. In short we find that, after appropriate rescaling of
the fields, \begin{equation}
\label{Gamma INV bosons}
\Gamma ^{-1}_{l,m,l',m'}\left( \mathbf{q},i\omega _{\nu }\right) =\left[ -i\omega _{\nu }-\mu ^{b}_{l}\left( \beta ,\mu \right) \right] \delta _{l,l'}\delta _{m,m'}+\sum _{i,j=x,y,z}\frac{\hbar ^{2}\mathbf{q}_{i}\mathbf{q}_{j}}{2m_{l,m,l',m'}^{b,ij}\left( \beta ,\mu \right) }.
\end{equation}
Thus the action \( S_{eff}^{(2)}\left[ \Delta ^{*},\Delta \right]  \)
describes an ideal Bose gas made out of bosons that propagate with
some effective mass \( m_{l,m,l',m'}^{b,ij}\left( \beta ,\mu \right)  \)
and chemical potential \( \mu ^{b}_{l}\left( \beta ,\mu \right)  \)
(these quantities are given by explicit formulae which we shall omit,
for brevity). In general the effective masses \( m_{l,m,l',m'}^{b,ij}\left( \beta ,\mu \right)  \)
represent anisotropic dispersion relations that are different for
PP in distinct internal states. Moreover, their off-diagonal values
(\( l\neq l',m\neq m' \)) may be finite, reflecting hybridization
between such states. Nevertheless in the BE limit we have\[
m_{l,m,l',m'}^{b,ij}\left( \beta ,\mu \right) \rightarrow \left\{ \begin{array}{l}
2m^{*}\textrm{ if }l=l'\textrm{ and }m=m'\\
\infty \textrm{ otherwise}
\end{array}\right. \]
indicating that the tightly-bound PP propagate freely as particles
of mass \( 2m^{*} \) without changing their angular momentum. This
simplifies (\ref{Sb}) to \begin{equation}
\label{Sb simpl}
S_{eff}^{(2)}\left[ \Delta ^{*},\Delta \right] =\beta \sum _{\mathbf{q}\nu }\sum _{l,m}\Delta ^{*}_{l,m,\mathbf{q}}\left( \omega _{\nu }\right) \left( -i\omega _{\nu }+\frac{\hbar ^{2}\left| \mathbf{q}\right| ^{2}}{4m^{*}}-\mu _{l}^{b}\left( \beta ,\mu \right) \right) \Delta _{l,m,\mathbf{q}}\left( \omega _{\nu }\right) 
\end{equation}
in that limit. The effective chemical potentials \( \mu _{l}^{b}\left( \beta ,\mu \right)  \),
on the other hand, are different for bosons with different values
of the angular momentum quantum number \( l \). The condition for
an instability of the gas of preformed pairs to a superconducting
state with angular momentum quantum number \( l \), Eq.~(\ref{Tc eq}),
corresponds to the BE condensation of the corresponding PP: \( \mu _{l}^{b}\left( \beta _{c},\mu _{c}\right) =0 \).
Typically, other PP with angular momentum quantum number \( l'\neq l \)
are also present in the normal state, just above \( T_{c} \), however
their chemical potential is \( \mu _{l'}^{b}\left( \beta _{c},\mu _{c}\right) <0 \)
at the transition so, unless they are also close to their own critical
temperature, they are only present in small number. One can thus neglect
such additional PP, and check the assumption a posteriori by ensuring
that the critical temperatures are quite different. Thus we eliminate
the sum on \( l \) from Eq.~(\ref{Sb simpl}). We can now very easily
deduce the explicit form of the density equation (\ref{gaussian-density-eq})
in this case. At the critical temperature, it is \begin{equation}
\label{density BE crucial}
n\left( \beta _{c},\mu \right) =\sum _{m=-l}^{l}\delta n_{l,m}\left( \beta _{c}\right) ,
\end{equation}
where \( \delta n_{l,m}\left( \beta _{c}\right) =2L^{-3}\sum _{\mathbf{q}}g\left( \beta _{c}\hbar ^{2}\left| \mathbf{q}\right| ^{2}/4m^{*}\right)  \)
is the density of a Bose gas that is exactly at its BE condsation
temperature, \( T_{c} \) (\( g\left( x\right) \equiv \left( e^{x}-1\right) ^{-1} \)
is the Bose distribution function). Evidently the fact that there
are \( 2l+1 \) values of \( m \) lowers the critical temperature:
it is given by \begin{equation}
\label{TcBE}
k_{B}T_{c}=3.315\frac{\hbar ^{2}}{2m^{*}}\left[ \frac{n}{2\left( 2l+1\right) }\right] ^{2/3}
\end{equation}
 i.e. it is the BE condensation temperature for \( n/[2(2l+1)] \)
bosons of mass \( 2m^{*} \) each. This reduces to the usual result
\cite{Nozieres-SchmittRink-85,Zwerger-92,SaDeMelo-Randeria-Engelbrecht-93,Zwerger-97}
only for bosons with internal angular momentum \( l=0 \): \begin{equation}
\label{TcBE for l eq 0}
k_{B}T_{c}=3.315\frac{\hbar ^{2}}{2m^{*}}\left( \frac{n}{2}\right) ^{2/3}\textrm{ for }l=0
\end{equation}
 For higher values of the angular momentum, the bosons still condense
all at the same temperature, but they do so as \( 2l+1 \) independent
Bose gases, each one corresponding to one of the degenerate angular
momentum states consistent with that value of \( l \). Obviously
the crucial consequence of this is that \emph{the supeconducting instability
in the low-density, strong-coupling limit always corresponds to the
BE codensation of pairs with \( l=0 \).} Evidently this result is
quite generic since the degeneracy in \( m \) is an unavoidable featue
of any central potential.%
\footnote{In contrast, the anisotropic effective interaction considered in \cite{Zwerger-97}
has, by definition, a single, non-degenerate bound state. Hence the
result that the BE limit of the critical temperature is given by (\ref{TcBE for l eq 0}),
in spite of the \( d \)-wave nature of the superconducting state.
}

\section{Conclusion\label{SEC-conclusion}}

In these pages we have revisited the old problem \cite{Balian-64,Anderson-Brinkman-74}
of exotic pairing via a central potential. According to a well-known
argument, a central potential can lead to a supercondcuting ground
state in which Cooper pairing takes place with a finite value of the
angular momentum quantum number \( l>0 \). We have demonstrated that
the natural framework for that to happen is provided by an interaction
in which the distance between the paired electrons is {}``locked''
to some finite value \( r_{0} \). We have then used the well-known
functional integral formulation of the BCS to Bose crossover \cite{Randeria-95}
to explore the behaviour of such models away from the original weak-coupling
limit. We have found evidence that a \emph{quantum phase transition},
in which the symmetry of the superconducting order parameter changes,
is associated, quite generally, with this type of interactions. The
phase trasition may occur at weak coupling as the density is varied,
making the preferred pairing channel change, or on approaching the
BE limit, where rotational symmetry is always restored. The later
is a consequence of very elementary energetic cosiderations, related
to the fact that the two-particle bound state with \( l=0 \) always
has lower energy than those with finite angular momentum quantum number.
Finally, we have discussed the BE limit of the critical temperature.
By neglecting preformed pairs with all but one value of \( l \),
we have been able to estimate the BE limit of \( T_{c} \) for different
values of the angular momentum quantum number. For \( l>0 \), we
have found the surprising result that \( T_{c} \) is not given by
the usual, simple formula for BE condensation, but instead it is considerably
suppressed due to the \( 2l+1 \) degeneracy of the corresponding
bound state. This may be an interesting example of the effect of internal
degrees of freedom of the constitutent bosons on the properties of
BE condensates, as discussed by Nozi\`eres \cite{Nozieres-95}.

Our remarks may serve as preliminary steps for a systematic exploration
of the possibility of rotational symmetry breaking in the possible
BCS state of degenerate Fermi gases \cite{Bruun-Castin-Dum-Burnett-99,Combescot-01,Holland-Kokkelmans-Chiofalo-Walser-01,Ohashi-Griffin-02},
where the distance \( r_{0} \) could be related to the shape of the
interatomic interaction. They also provide an indication that some
of the features that we have identified
\cite{Quintanilla-Gyorffy-00,Quintanilla-Gyorffy-02,Quintanilla-Gyorffy-Annett-Wallington-02}
in the new, delta shell model may be relevant to a larger, and important,
class of interaction potentials.

\section*{Acknoldegments}

We thank James F. Annett and Jonathan P. Wallington for many useful
discussions on this topic. JQ acknowledges financial support from
FAPESP (Brazil; process No.~01/1046 1-8).

\small

\bibliographystyle{apsrev}
\bibliography{bibliography}

\end{document}